\documentclass[pdflatex,sn-mathphys-num]{sn-jnl}

\usepackage{graphicx}%
\usepackage{multirow}%
\usepackage{amsmath,amssymb,amsfonts}%
\usepackage{amsthm}%
\usepackage{mathrsfs}%
\usepackage[title]{appendix}%
\usepackage{xcolor}%
\usepackage{textcomp}%
\usepackage{manyfoot}%
\usepackage{booktabs}%
\usepackage{algorithm}%
\usepackage{algorithmicx}%
\usepackage{algpseudocode}%
\usepackage{listings}%
\usepackage[T1]{fontenc}
\usepackage[utf8]{inputenc}
\usepackage{tabularx}
\usepackage{xcolor}
\usepackage{comment}
\usepackage{soul}

\newcommand{\revision}[1]{{#1}}

\begin{document}

\title[Article Title]{Time-Reversal and Reversible Dynamics in Cavity QED for Quantum Metrology}

\author*[1]{\fnm{Colombo} \sur{Simone}}\email{simone.colombo@uconn.edu}
\equalcont{These authors contributed equally to this work.}

\author*[2]{\fnm{Pedrozo-Pe\~nafiel} \sur{Edwin}}\email{epedrozo@ufl.edu}
\equalcont{These authors contributed equally to this work.}

\affil*[1]{\orgdiv{Department of Physics}, \orgname{University of Connecticut}, \orgaddress{\street{196 Auditorium Road}, \city{Storrs}, \postcode{06269-3046}, \state{Connecticut}, \country{USA}}}

\affil*[2]{\orgdiv{Department of Physics}, \orgname{University of Florida}, \orgaddress{\street{2001 Museum Rd}, \city{Gainesville}, \postcode{32611-8440}, \state{Florida}, \country{USA}}}


\abstract{Quantum-enhanced metrology relies on entanglement to achieve sensitivities beyond the standard quantum limit. While remarkable progress has been made in generating highly entangled many-body states, extracting their metrological advantage remains a central challenge because the encoded information is often inaccessible to realistic measurements. A key development of the past decade has been the realization that many-body interactions can play a dual role: they can be used not only to generate entanglement, but also to decode it. This idea underlies interaction-based readout and \revision{operational} time-reversal protocols, in which controlled nonlinear dynamics transform weakly encoded signals into experimentally accessible observables. Cavity quantum electrodynamics (QED) provides a particularly powerful setting for these approaches because it combines collective enhancement, tunable interactions, and controllable reversibility within a single platform. In this review, we discuss the emergence of time-reversal protocols in cavity QED, from their conceptual roots in Loschmidt echoes to modern implementations of signal amplification through a time-reversed interaction (SATIN), \revision{quantum-phase magnification}, and more general interaction-based readout schemes. 
We examine the physical mechanisms that enable reversible many-body dynamics, review key experimental demonstrations, and discuss future directions involving complex entangled states, nonlinear decoding, and emerging quantum platforms. Together, these developments suggest that the ability to decode quantum information may become as important as the ability to generate it, establishing reversible many-body dynamics as a central resource for quantum-enhanced sensing.
}




\maketitle

\section{Introduction}\label{sec1}
Quantum metrology exploits uniquely quantum resources to improve the precision of measurements beyond the standard quantum limit (SQL), which arises from the projection noise of \revision{$N$} independent particles \revision{and corresponds to a phase uncertainty $\Delta\phi_\mathrm{SQL}=N^{-1/2}$}~\cite{wineland1992,wineland1994,giovannetti2004quantum,toth2014quantum}. Over the past several decades, a broad range of entangled states, including spin-squeezed states, Dicke states, Schr\"{o}dinger-cat states, and more general non-Gaussian many-body states, have been proposed and realized as resources for enhanced sensing~\cite{esteve2008squeezing,gross2010nonlinear,hamley2012spin,ma2011quantum,pezze2018quantum,Huang_review_2024,bao2020spin,kaubruegger2025progress}. Their use promises substantial gains in applications ranging from atomic clocks~\cite{ludlow2015review,gil2014,pedrozo2020entanglement,schulte2020prospects,colombo2022entanglement,leibfried2004,nichol2022elementary,dietze2026,robinson2024direct,yang2025clock,kaubruegger2025progress,eckner2023realizing,cao2024multi} and magnetometers to inertial sensors and tests of fundamental physics~\cite{pezze2018quantum,szigeti2021improving,greve2022entanglement,demille2024quantum,terrano2022comagnetometer,budker2007optical,degen2017quantum,ye2024essay}. 

Despite remarkable advances in generating entangled states, their practical deployment in quantum sensors faces an important challenge: extracting the metrological advantage encoded in fragile many-body correlations~\cite{Str14b,Fro15,Col21}. In principle, highly entangled states can approach Heisenberg-limited sensitivity~\cite{pezze2018quantum,Huang_review_2024,kaubruegger2025progress,Dav16,cao2024multi,finkelstein2024universal}\revision{, where the parameter uncertainty is $\Delta\phi=N^{-1}$}. In practice, however, realizing this advantage often requires detection capabilities capable of resolving fluctuations below the SQL, a requirement that becomes increasingly demanding as system size grows~\cite{Hos16b,Nol17,Dav16}. Consequently, the development of robust measurement strategies has emerged as one of the central problems of quantum-enhanced metrology~\cite{Nol17,Hai18,Sch19,Col21}. 

A particularly elegant solution to this challenge exploits many-body interactions not only to generate entanglement but also to decode it~\cite{Dav16,Hos16b}. In these protocols, the same nonlinear dynamics responsible for creating a quantum-enhanced probe are subsequently used to transform the encoded information into observables accessible with realistic detection~\cite{Dav16,Col21}. This concept, broadly known as interaction-based readout, has significantly expanded the class of entangled states that can be exploited experimentally and has reduced the stringent requirements traditionally associated with quantum-limited detection~\cite{Nol17,Hai18,Sch19,Col21}. 

Among interaction-based protocols, time-reversal strategies occupy a special position~\cite{Mac16,Dav16}. Their conceptual roots can be traced back to Loschmidt echoes, originally introduced to investigate the reversibility of many-body dynamics and the sensitivity of quantum evolution to perturbations~\cite{Gou12}. In those settings, the degree to which an initially prepared state could be reconstructed after forward and backward evolution provided insight into decoherence, scrambling, and the emergence of irreversibility in complex quantum systems~\cite{Gou12,Li22c}. More recently, however, similar inversion procedures have acquired a fundamentally different role~\cite{Mac16,Dav16}. Rather than diagnosing reversibility itself, reversed dynamics can be used as a metrological resource~\cite{Mac16,Dav16}. Following the proposal of twisting echoes by Davis \emph{et al.}~\cite{Dav16}, several experiments demonstrated that inverse nonlinear evolution can amplify otherwise inaccessible quantum signatures into collective observables detectable with finite-resolution measurements~\cite{Lin16,Col21,Mao22,Li22c,Gil21}. \revision{Here, time reversal refers to implementing the inverse propagator $U^\dagger$, typically through a controlled reversal of the sign of an engineered Hamiltonian, $\hat H\rightarrow -\hat H$. The fundamental time-reversal operation is an antiunitary transformation.}

This approach has been implemented using a variety of physical platforms, including spinor Bose--Einstein condensates, trapped ions, and cavity QED systems~\cite{Lin16,Gil21,Col21,Mao22}. Particularly notable is the cavity-based SATIN protocol (signal amplification through a time-reversed interaction), in which cavity-mediated one-axis twisting is dynamically reversed to transform weak encoded phases into large collective spin displacements while preserving near-Heisenberg sensitivity~\cite{Col21}. At the same time, it has become increasingly clear that exact time reversal is only one member of a broader family of nonlinear decoding protocols~\cite{Hos16b,Nol17,Sch19,Li22c}. The general interaction-based-readout framework recognizes that the final operation need not perfectly invert the state-preparation dynamics. Instead, it should be chosen to optimize the extraction of information under realistic experimental constraints, including finite detection resolution, decoherence, and technical noise~\cite{Nol17,Hai18,Sch19}. 

Cavity QED provides an especially attractive setting in which to explore these ideas. Optical cavities enable the generation of entanglement through both dissipative and coherent mechanisms~\cite{Li21b,Hos16c,Cox16b,Sch10b,Ler10,Bra19}, provide exquisite control over collective interactions~\cite{Li21b,Sch10b,Bra19}, and offer the possibility of dynamically reversing effective Hamiltonians through experimentally accessible parameters such as laser detuning~\cite{Col21}. \revision{For an open system, this inversion applies to the coherent Hamiltonian dynamics. Cavity leakage, spontaneous emission, technical noise, and information transferred to the environment continue to evolve forward and limit the fidelity of the reversal.} These capabilities have positioned cavity QED at the forefront of recent developments in metrological time reversal~\cite{Col21,Li22c,Zap25}. 

In this review, we examine the emergence of time-reversal protocols in cavity QED and their applications to quantum-enhanced sensing~\cite{Col21,Li22c,Zap25}. \revision{The review is centered on cavity QED and spin–boson interactions, with selected examples from other platforms included to highlight broader conceptual connections.} We first discuss the mechanisms through which cavity QED generates entanglement and highlight the distinction between dissipative and coherent interactions~\cite{Li21b,Hos16c,Cox16b,Sch10b,Ler10,Bra19}. We then review the conceptual development from Loschmidt echoes to SATIN and interaction-based readout, emphasizing the differences between reversibility as an object of study and reversibility as a metrological tool~\cite{Gou12,Mac16,Dav16,Nol17,Col21}. Finally, we discuss recent experimental advances and outline future directions, arguing that reversible many-body dynamics may constitute a central paradigm for the next generation of quantum sensors~\cite{Col21,Li22c,Zap25}.

\section{Entanglement generation in cavity QED}
Cavity QED provides a natural setting for generating entanglement in large atomic ensembles because many particles couple collectively to a common optical mode~\cite{tanji2011interaction}. This collective coupling allows optical fields to mediate effective spin-spin interactions, enhance measurement backaction, and convert microscopic quantum fluctuations into observable optical signals~\cite{Li21b,tanji2011interaction}. For the purposes of this review, the important point is not merely that cavities generate entanglement, but that they can do so through mechanisms with very different implications for reversibility~\cite{Col21,Li22c}. Broadly, cavity-based entanglement generation falls into two categories. In the first, information about the collective atomic state is extracted through the cavity output field. The resulting entanglement is measurement-induced and therefore dissipative~\cite{kuzmich1998atomic,appel2009mesoscopic,saffman2009spin,Hos16c,Cox16b}. In the second, the cavity mediates coherent many-body interactions that generate entanglement through approximately unitary dynamics~\cite{Sch10b,Ler10,Bra19,Li21b,Hos16b}. This distinction is central for time-reversal metrology: measurement-induced entanglement can be extremely useful for quantum enhancement, but only coherent dynamics can be \revision{perfectly} inverted and used as a nonlinear decoding operation~\cite{Li21b,Col21}.

\subsection{Dissipative generation of entanglement}
One of the most successful approaches to entanglement generation in cavity QED is based on quantum non-demolition (QND) measurement of collective spin observables~\cite{kuzmich1998atomic,appel2009mesoscopic,Hos16c,Cox16b,chaparro2026dissipative}. In a typical dispersive implementation, an ensemble of effective spin-$1/2$ particles shifts the resonance frequency of an optical cavity by an amount proportional to a collective spin projection, such as $S_z$~\cite{appel2009mesoscopic,Takano09,Hos16c,Cox16b}. The phase or intensity of the transmitted or reflected light therefore carries information about $S_z$. Detecting this light reduces the uncertainty of the collective spin projection and conditionally prepares a spin-squeezed state~\cite{appel2009mesoscopic,Hos16c,Cox16b}. This measurement-based route has produced some of the largest metrological gains in atomic ensembles and has enabled clocks and interferometers operating beyond the standard quantum limit~\cite{Hos16c,Cox16b,robinson2024direct,yang2025clock,greve2022entanglement}. Its strength is robustness: the entanglement is generated by observing the system, and feedback or feedforward can convert the conditional state into a practical metrological resource~\cite{Cox16b}. However, from the perspective of time reversal, QND squeezing has an important limitation. The measurement process exports information about the atomic state into the environment through the cavity output field~\cite{Li21b}. Once this information has leaked out, the many-body evolution cannot in general be exactly undone by a subsequent unitary operation on the atoms alone. Thus, measurement-induced entanglement is powerful for quantum enhancement, but it is not the natural starting point for Hamiltonian-level time reversal~\cite{Li21b,Col21}.

\subsection{Unitary cavity-mediated interactions as a platform for reversible metrology}

Cavity QED can also generate entanglement through coherent interactions rather than information extraction. In this regime, the cavity field mediates an effective nonlinear Hamiltonian for the collective spin. The paradigmatic example is one-axis twisting (OAT), introduced by Kitagawa and Ueda~\cite{kitagawa1993}, with Hamiltonian
\begin{equation}
\hat H_{\mathrm{OAT}}=\hbar\chi \hat S_z^2,\label{eq:OAT}
\end{equation}
where $\chi$ is the effective shearing rate.

\begin{figure}[th!]
    \centering
    \includegraphics[width=0.35\linewidth]{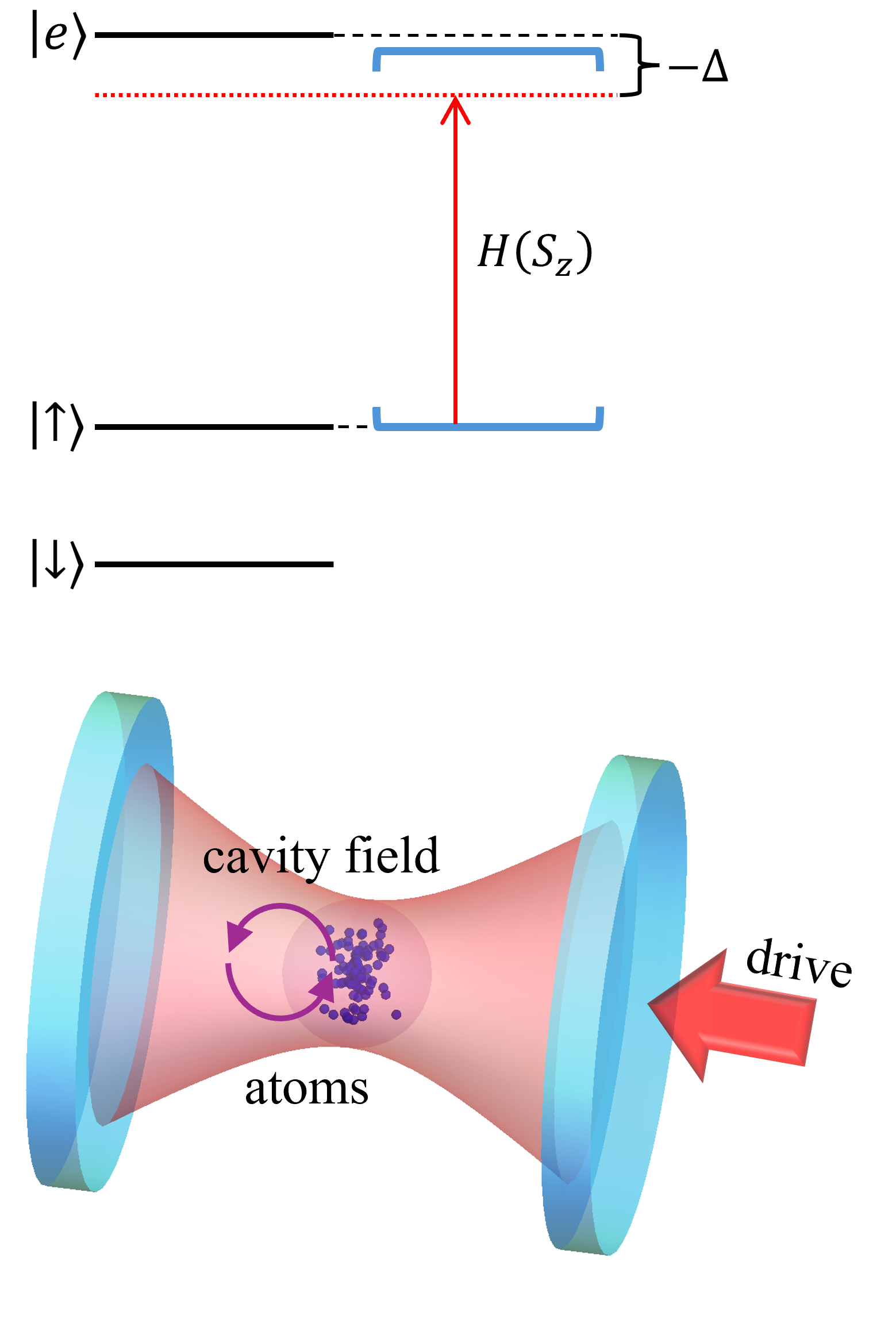}
    \caption{\textbf{Cavity-mediated one-axis twisting as a source of reversible many-body entanglement.} \revision{An atomic ensemble coupled to an optical cavity realizes the effective Hamiltonian $\hat H_{\mathrm{OAT}}=\hbar\chi\hat S_z^2$. The resulting nonlinear spin dynamics generate squeezing at short times and non-Gaussian entangled states at longer times. The predominantly coherent interaction and its controllable sign make this system suitable for time-reversal and interaction-based-readout protocols. See the main text for details of the cavity-feedback mechanism.}}
    \label{fig:generalCavitySqueezing}
\end{figure}

The physical mechanism underlying cavity-mediated one-axis twisting is illustrated in Fig.~\ref{fig:generalCavitySqueezing}. In cavity-feedback squeezing~\cite{Sch10b,Ler10,Bra19}, fluctuations of the collective spin shift the cavity resonance, modifying the intracavity photon number. The resulting ac Stark shift acts back on the atoms, producing a spin precession rate that depends on $S_z$~\cite{Sch10b,Ler10,Bra19}. At short evolution times, this nonlinear shearing transforms a coherent spin state into a Gaussian spin-squeezed state. At longer times, the state enters the over-squeezed regime, where the distribution wraps around the Bloch sphere and develops strongly non-Gaussian structure~\cite{Str14b,Col21,Li22c,Dav16}. Such states may contain large quantum Fisher information even when conventional squeezing parameters no longer indicate useful metrological gain~\cite{Str14b,Dav16,pezze2018quantum}.

\revision{The strength of this interaction is set by a few experimentally accessible parameters.
For $N$ atoms coupled to a cavity of linewidth $\kappa$ with single-atom coupling $g$ and excited-state linewidth $\Gamma$, the relevant quantities are the single-atom cooperativity $\eta=4g^2/(\kappa\Gamma)$ and the normalized atomic and cavity detunings $x_a=2(\omega_l-\omega_a)/\Gamma$ and $x_c=2(\omega_l-\omega_c)/\kappa$ of the probe laser at frequency $\omega_l$, with dispersive and absorptive atomic responses $\mathcal{L}_d(x_a)=-x_a/(1+x_a^2)$ and $\mathcal{L}_a(x_a)=1/(1+x_a^2)$.
We consider the atoms prepared in a coherent spin state on the equator of the generalized Bloch sphere, i.e., $\langle\hat S_z\rangle=0$ and $\langle N_\uparrow\rangle=N/2$.
Eliminating the cavity field and the excited state in the dispersive, weakly saturated regime,
the photon-number-dependent light shift yields an effective Hamiltonian of the form of Eq.~\eqref{eq:OAT},
with
\begin{equation}
  \chi = -n_0\,\eta^2\,\mathcal{T}_0\mathcal{L}_a(x_a)\mathcal{L}_d(x_a)\left(\,1-x_c x_a+ \frac{N}{2}\eta\right),
\label{eq:chi}
\end{equation}
where $\mathcal{T}_0=
\left[
\left(
1+\frac{N\eta}{2}\mathcal L_a(x_a)
\right)^2
+
\left(
x_c+\frac{N\eta}{2}\mathcal L_d(x_a)
\right)^2
\right]^{-1}\nonumber
$ is the normalized cavity transmission, $n_0$ the intracavity photon-flux~\cite{Li21b}.
The detunings thus set both the magnitude and the sign of $\chi$, the coupling and linewidths enter through $\eta$,
and the drive power sets the intracavity photon flux $n_0$, and hence the accumulated shear $\propto n_0$.
The sign of $\chi$ can be reversed in two ways.
For $\omega_c=\omega_a$, the drive can be detuned to the opposite side of the atomic and cavity resonances,
$(x_a,x_c)\to(-x_a,-x_c)$, resulting in a
$\chi\to-\chi$ change; here, the dissipation rates discussed below are unchanged.
Alternatively, the drive can be detuned to the other side of the dressed cavity resonance without changing the sign of
the detuning from $\omega_a$, which reverses the slope of the intracavity photon number with respect to $\hat S_z$
and hence the sign of $\chi$; in this case the dissipation rates are in general different in the two configurations.
In practice, this is done by switching the drive frequency~\cite{Col21},
together with reversing the phase of the radio-frequency drive in twist-and-turn dynamics, where nonlinear one-axis twisting competes with a transverse collective-spin drive~\cite{Li22c}.}

Unlike measurement-induced squeezing, cavity-mediated OAT can operate close to the unitary limit~\cite{Bra19}. Although practical implementations remain subject to photon loss and spontaneous emission, the entanglement is generated predominantly through coherent many-body evolution~\cite{Li21b,Bra19}. This distinction has important consequences for quantum metrology. In several cavity implementations, the sign of the effective interaction can be controlled through experimentally accessible parameters such as the optical detuning~\cite{Dav16,Col21,Li21b}, allowing an entangling evolution to be followed by an inverse or approximately inverse operation.

Rather than serving solely as a mechanism for state preparation, cavity-mediated interactions can therefore be used to manipulate quantum correlations after signal encoding~\cite{Dav16,Nol17,Col21}. This additional level of control enables protocols in which the same dynamics responsible for generating entanglement are later exploited to transform information encoded in complex many-body states into collective observables accessible with realistic detection~\cite{Dav16,Col21}. It is this capability that underlies the time-reversal and interaction-based readout protocols discussed in the following sections.

\revision{\paragraph{Dissipation.}
Cavity-mediated twisting is accompanied by two independent decoherence channels: collective dephasing
from cavity photons that carry away information about $\hat S_z$, and single-atom decoherence from free-space scattering.
The dynamics is described by the effective master equation
\begin{equation}
  \frac{\partial\hat\rho}{\partial t}
  = -i\bigl[\chi\hat S_z^2,\hat\rho\bigr]
    +\Gamma_\phi\,\mathcal{D}[\hat S_z]\hat\rho
    +\gamma_{\rm sc}\sum_{i=1}^{N}\mathcal{D}[\hat\sigma_z^{(i)}]\hat\rho,
\label{eq:lindblad}
\end{equation}
with $\hat{\rho}$ the density operator, $\mathcal{D}[\hat L]\hat\rho=\hat L\hat\rho\hat L^\dagger-\tfrac12\{\hat L^\dagger\hat L,\hat\rho\}$ the jump operator,
which separates the reversible twisting from collective dephasing at rate $\Gamma_\phi$ and free-space scattering at rate $\gamma_{\rm sc}=\eta\mathcal{L}_a(x_a)$.
The explicit jump operators depend on the level scheme; the form above is the simplest model of Rayleigh scattering.
Both rates vanish in the dispersive limit $|x_a|\gg1$ faster than $\chi$, which together with the collective enhancement $N\eta$
of the twisting explains why large cooperativity and large detuning enable near-unitary squeezing~\cite{Bra19}. Because $\hat H$ and the dephasing operator are both diagonal in the Dicke basis, the dephasing channel suppresses coherences between states separated by $|m_k-m_l|$ by the Gaussian factor $\exp[-\tfrac12(m_k-m_l)^2\Gamma_\phi n_0 t]$, while scattering contracts the Bloch vector by $e^{-2\gamma_{\rm sc}n_0 t}$ and the quantum Fisher information by its square.
In time-reversal protocols, this decoherence acts during both the twisting and untwisting stages, which is why it ultimately limits the achievable metrological gain.}

\section{Time reversal, SATIN, and interaction-based readout}

The term time reversal is used in several related but distinct ways. In the Loschmidt-echo tradition, the reversal is itself the object of study. One evolves a system forward, attempts to invert the dynamics, and asks how much of the initial state is recovered~\cite{Gou12,Mor02,Meu05,wang2026quantum}. The echo then quantifies sensitivity to perturbations, decoherence, and imperfect control.

\begin{figure}[h]
    \centering
    \includegraphics[width=1.\linewidth]{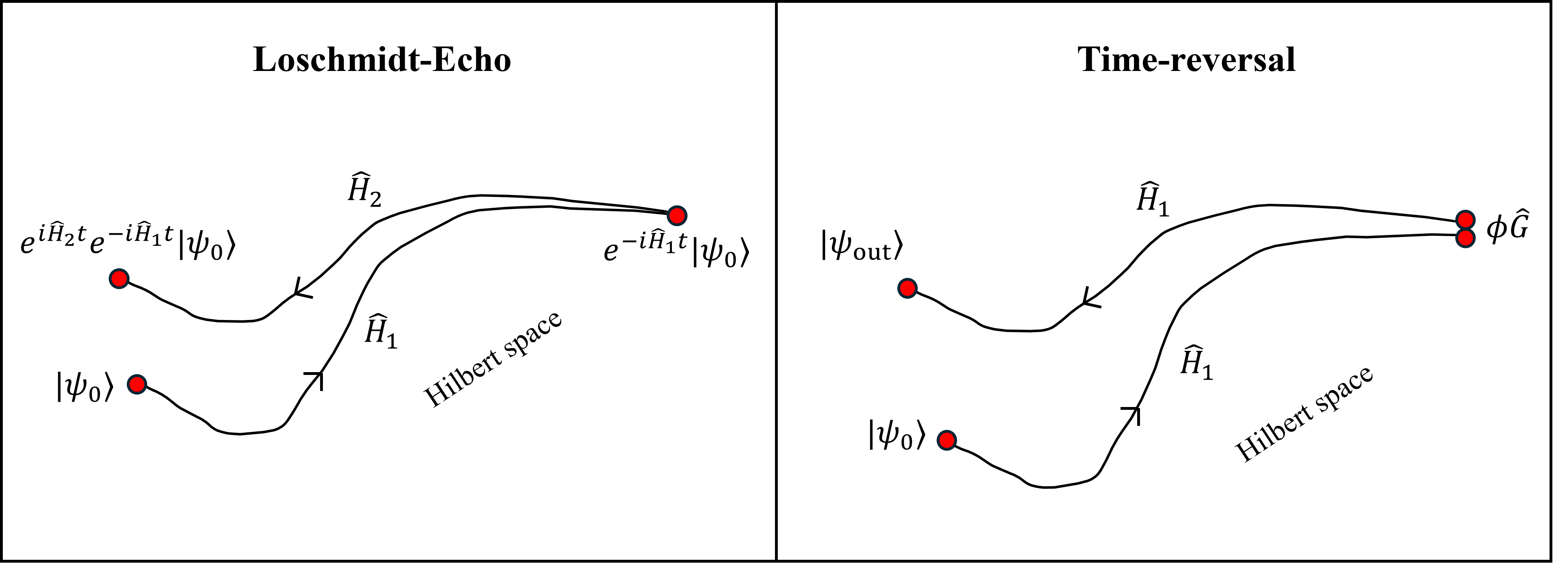}
    \caption{\textbf{Conceptual distinction between Loschmidt echoes and metrological time-reversal protocols.}
In a Loschmidt echo (left), the objective is to reverse a quantum evolution and quantify the fidelity of state recovery, thereby probing reversibility, decoherence, and sensitivity to perturbations. In metrological time reversal (right), the inverse evolution serves a different purpose: it transforms information encoded in an entangled many-body state into a measurable collective observable. Reversibility is therefore not the quantity being measured but the resource that enables enhanced readout.}
    \label{fig:losch_SATIN_comp}
\end{figure}

In quantum metrology, the same formal structure can play a different role. The backward evolution is no longer used primarily to test reversibility, but to read out a signal. A weak phase is encoded between two nonlinear operations, and the second operation converts the encoded information into an observable compatible with realistic detection~\cite{Dav16,anders2018phase,Hos16b,Nol17,Col21}. Cavity QED is particularly well suited for illustrating this distinction because it contains both limits: literal reversal of Jaynes--Cummings dynamics in early echo work~\cite{Mor02,Meu05} and engineered sign reversal of cavity-mediated one-axis twisting in SATIN~\cite{Hos16b,Li21b,Col21}. SATIN provides one of the most direct realizations of metrological time reversal in cavity QED: the sign of the cavity-mediated interaction is reversed so that the many-body dynamics act as an untwisting operation after signal encoding. The conceptual distinction between these two uses of reversed dynamics is summarized in Fig.~\ref{fig:losch_SATIN_comp}.

These developments motivate a distinction between three related but conceptually different frameworks. Loschmidt echoes probe reversibility. SATIN uses Hamiltonian reversal as a metrological resource. Interaction-based readout is the broader framework in which the final interaction is chosen to extract information efficiently, whether or not it is an exact inverse~\cite{Nol17,Hai18,Mir18,Sch19,Baa22}.

\subsection{Loschmidt echoes: reversibility as an object of study}

A Loschmidt echo starts from an initial state $|\psi_0\rangle$, evolves it forward under one Hamiltonian, and then attempts to reverse the dynamics using a second evolution. The standard figure of merit is the return probability

\begin{equation}
M(t)=\left|\langle \psi_0|e^{\frac{i}{\hbar} \hat H_2 t}e^{-\frac{i}{\hbar} \hat H_1 t}|\psi_0\rangle\right|^2,
\end{equation}

which quantifies the overlap between the initial state and the state obtained after the forward and backward evolutions~\cite{Gou12}. In the ideal case of perfect reversal, $M(t)=1$. In realistic systems, imperfections, perturbations, and decoherence reduce the return probability, making the Loschmidt echo a probe of sensitivity to perturbations and the emergence of irreversibility~\cite{Gou12}.

This is the sense in which Loschmidt echoes have traditionally been used in many-body physics and quantum chaos. The central question is not how to estimate an unknown parameter with maximal precision, but how faithfully a quantum evolution can be reversed. Early cavity-QED experiments followed this logic: reversing Jaynes--Cummings dynamics allowed the recovery of mesoscopic field states to be used as a diagnostic of coherence and decoherence in the cavity field~\cite{Mor02,Meu05}. What is measured is not an amplified estimate of a weak signal, but the success of the inversion itself.

The connection between echo protocols and quantum metrology emerged later. In~\cite{Mac16}, an entangling operation, a phase encoding step, and an inverse evolution were combined so that the return probability to the input state provides direct access to the quantum Fisher information. This protocol imports echo concepts into metrology, but its primary observable remains a fidelity or overlap. It is therefore useful to distinguish such approaches from SATIN and related interaction-based readout protocols, where the goal is not to measure reversibility itself, but to use reversed dynamics to transform a weak encoded phase into a robust collective signal. In the latter case, reversibility is not the quantity being measured; it is the resource that enables the measurement.

\subsection{SATIN: reversibility as a metrological resource}

The role of time reversal changes fundamentally when the objective is parameter estimation rather than the characterization of reversibility itself. In this setting, the inverse evolution is not used to determine whether a state can be recovered, but to transform information encoded in an entangled state into a measurable collective observable. A generic metrological time-reversal protocol has the form

\begin{equation}
|\psi_{\mathrm{out}}\rangle
=
U^{\dagger}
e^{-i\phi \hat G}
U
|\psi_0\rangle,
\end{equation}

where $U$ generates entanglement, $e^{-i\phi \hat G}$ encodes the parameter of interest, and the inverse evolution $U^\dagger$ acts as a decoding operation.

This idea was introduced in the twisting-echo proposal of Davis \textit{et al.}~\cite{Dav16}. The central observation was that the inverse interaction can amplify the measurement signal while simultaneously relaxing the requirements on detection resolution. Rather than attempting to directly resolve the fine structure of an entangled many-body state, the protocol converts the encoded phase into a macroscopic collective displacement that remains observable even with detection noise at the coherent-spin-state level. The purpose of the inverse evolution is therefore not to erase entanglement, but to make its metrological content experimentally accessible.

\begin{figure}
    \centering
    \includegraphics[width=0.995\linewidth]{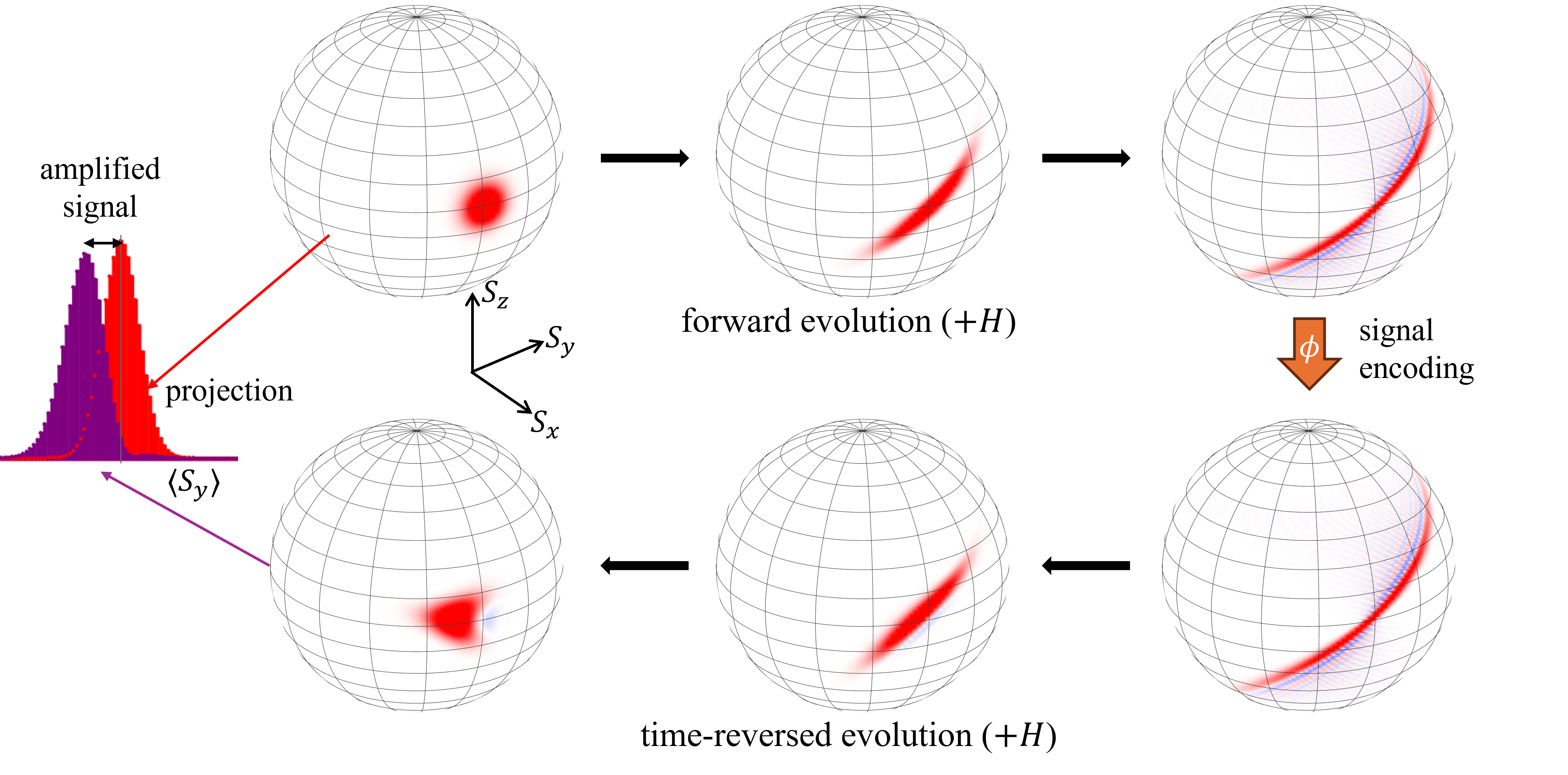}
    \caption{\textbf{Schematic of the SATIN (signal amplification through a time-reversed interaction) protocol.}
An initially coherent spin state undergoes nonlinear evolution under a one-axis-twisting Hamiltonian, generating an entangled state with non-Gaussian quantum correlations. A weak phase is then encoded as a small rotation of the collective spin. Subsequent evolution under the reversed interaction untwists the state and amplifies the encoded phase into a large collective-spin displacement. The final signal can be resolved with detection noise at or near the coherent-spin-state level, allowing near-Heisenberg-limited sensitivity without single-particle resolution.}
    \label{fig:timeReversal_general}
\end{figure}

SATIN (signal amplification through a time-reversed interaction) provides a direct realization of this idea in cavity QED~\cite{Col21}. In that experiment, cavity-mediated light shifts generate an effective one-axis-twisting Hamiltonian, and the sign of the interaction is reversed by changing the optical detuning~\cite{Li21b,Col21}. A first stage generates a non-Gaussian entangled state. After phase encoding, evolution under the opposite Hamiltonian untwists the state and maps the encoded phase onto a large displacement of a collective spin component that can be measured directly. The principle of the protocol is illustrated schematically in Fig.~\ref{fig:timeReversal_general}.

The conceptual importance of SATIN extends beyond its specific implementation. It demonstrates that the same many-body interactions responsible for generating entanglement can also be used to decode it. In this sense, SATIN represents one of the clearest examples of reversibility being employed as a metrological resource rather than as an object of study. The inverse dynamics are not used to probe irreversibility, as in a Loschmidt echo, but to enhance the accessibility of quantum information stored in highly entangled states.

\subsection{Interaction-based readout as the broader framework}

Although metrological time reversal has attracted considerable attention, exact inversion is not generally required to achieve optimal sensitivity. This observation motivates the broader concept of interaction-based readout (IBR), in which a second many-body operation is applied after phase encoding in order to make the encoded information accessible to the available measurement apparatus.

The general structure of an interaction-based readout protocol is

\begin{equation}
|\psi_\phi\rangle
=
U_2 e^{-i\phi \hat G} U_1 |\psi_0\rangle ,
\end{equation}

where $U_1$ prepares a nonclassical probe state and $U_2$ is chosen to optimize the final measurement~\cite{Nol17}. Exact time reversal corresponds to the special case $U_2=U_1^\dagger$, but the broader framework does not require this condition.

This viewpoint was formalized by Nolan \textit{et al.}~\cite{Nol17}, who identified general conditions under which interaction-based readout can saturate the quantum Cramér--Rao bound. For parity-structured protocols, optimal sensitivity can be achieved even when the final operation differs substantially from an exact inverse. The essential requirement is that the readout preserve the metrologically relevant information in a basis accessible to the measurement.

Subsequent work clarified that optimal readout generally does not require exact inversion. Instead, the decoding operation can be tailored to the dominant experimental limitations. 
In particular, interaction-based readout protocols can be designed to maximize robustness against detection noise~\cite{Hai18}, and may even outperform exact time reversal when the underlying dynamics differ from one-axis twisting~\cite{Mir18}. More generally, asymmetric untwisting protocols and generalized twisting echoes have identified regimes where partial or overcompensated reversal provides improved metrological performance~\cite{Sch19}. Analytical studies incorporating decoherence further showed that measurement-after-interaction strategies remain effective even when dissipation limits the amount of achievable squeezing~\cite{Baa22}.

SATIN naturally fits within the broader interaction-based-readout framework, with the particular feature that the decoding operation is implemented through a reversal of the entangling dynamics. More generally, interaction-based readout does not require exact inversion. The central idea is to use a second many-body operation to convert the encoded information into an experimentally accessible observable while preserving as much of the available quantum enhancement as possible.

\begin{table}[h]
\caption{Comparison of Protocols}\label{tab1}%
\centering
\small
\begin{tabular}{@{}p{0.12\textwidth}p{0.19\textwidth}p{0.19\textwidth}p{0.19\textwidth}p{0.17\textwidth}@{}}
\toprule
Protocol class & Post-encoding operation & Observable & Purpose & References \\
\midrule
Loschmidt echo & Attempted inversion of earlier dynamics & Return probability or fidelity & Probe reversibility, perturbation sensitivity, or decoherence & Conceptual precursor~\cite{Gou12,Mor02,Meu05} \\
Metrological time reversal & Exact or approximate inverse, often $U^{\dagger}$ or effective $-H$ & Amplified collective spin signal after untwisting & Convert a small encoded phase into an accessible readout & Pure time-reversal ~\cite{Dav16,Lew20,Col21} \\
Interaction-based readout & Any engineered readout interaction $U_2$, inverse or not & Measurement statistics in a convenient basis & Preserve sensitivity while improving robustness to detection noise & Broad umbrella frameworks~\cite{Nol17,Hai18,Mir18,Sch19,Baa22} \\
\bottomrule
\end{tabular}
\end{table}

\section{Advantages of nonlinear readout in spin metrology}

The conceptual distinctions between Loschmidt echo, metrological time reversal, and interaction-based readout were introduced in the previous section. The question addressed here is when a post-encoding interaction becomes necessary in quantum-enhanced metrology. Two limitations are particularly important. The first is intrinsic to the structure of the many-body state: the most sensitive states often lie beyond the Gaussian regime, where squeezing parameters and linear error propagation no longer provide a complete description of the available metrological gain. \revision{The metrological gain $\mathcal{G}$ is the squared improvement in the uncertainty estimation a quantity $\phi$ relative to the standard quantum limit (SQL),
\begin{equation}
\mathcal{G}
\equiv
\left(\frac{\Delta\phi_{\mathrm{SQL}}}{\Delta\phi}\right)^2,
\end{equation}
where $\Delta\phi$ is the uncertainty achieved by the measurement protocol. For a quantum state with quantum Fisher information $F_Q$, the quantum Cramér–Rao bound sets the attainable uncertainty $\Delta\phi \geq 1/\sqrt{F_Q}$. The metrological gain therefore depends both on the information encoded in the quantum state and on how effectively the measurement extracts that information.} The second is instrumental: finite detection resolution can erase that gain even when the underlying quantum Fisher information remains large. Although conceptually distinct, these limitations often emerge simultaneously in the same operating regime, motivating the use of nonlinear readout protocols~\cite{Str14b,Dav16,Hos16b,Col21}.

\subsection{Beyond Gaussian squeezing and Bloch-sphere curvature}

The standard squeezing description assumes that the collective state remains close to a coherent spin state, allowing the Bloch sphere to be linearized locally. In this regime, the metrological content is well captured by the first two moments of a transverse collective spin component, and the Wineland parameter as
\begin{equation}
   \xi_R^2 = \frac{N \revision{\mathrm{min}_{\mathbf{n}_\perp}}(\Delta \hat{S}_{\mathbf{n}_\perp})^2}{|\langle \hat{\mathbf{S}} \rangle|^2},
\end{equation}
provides a useful measure of spectroscopic enhancement. \revision{Here, $\mathbf{n}_\perp$ is a unit vector in the plane perpendicular to the mean collective-spin $\langle\hat{\mathbf{S}}\rangle$.} States with $\xi_R^2 < 1$ exhibit sensitivity beyond the standard quantum limit and can be understood within a Gaussian description of the collective spin fluctuations. 

This description breaks down once nonlinear dynamics drives the state into the over-squeezed regime. The quasiprobability distribution bends around the Bloch sphere, higher moments become important, and no single collective-spin observable provides a faithful description of the available metrological information~\cite{Str14b,Dav16,Hua18a,Hua18b,Col21}. Although the conventional squeezing parameter may no longer indicate useful enhancement, the state can still possess large quantum Fisher information and remain a valuable resource for precision measurements.

T\revision{h}is point was demonstrated experimentally by Strobel \emph{et al.}~\cite{Str14b}. In that work, non-Gaussian states generated by nonlinear evolution no longer satisfied the \revision{usual} squeezing criterion, yet their full probability distributions still contain metrological\revision{ly} useful information. The relevant condition was not simply $\xi_R^2 < 1$, but rather $F_Q > N$, which certifies multiparticle entanglement useful for sub-standard-quantum-limit sensing. The phase information was encoded in the complete probability distribution rather than in the mean and variance of a single collective observable. 

A compact way to express this idea is through the Hellinger distance between probability distributions. Defining $
d_H^2(\phi) = \tfrac{1}{2}\sum_z \left(\sqrt{P_z(\phi)}-\sqrt{P_z(0)}\right)^2,
$ one finds for small phase shifts, 
\revision{$d_H^2(\theta)=\frac{F_C}{8}\theta^2+O(\theta^3)$,
where $F_C$ is the classical Fisher information associated with the chosen measurement. 
The curvature of the Hellinger distance therefore quantifies the information accessible through that measurement. The quantum Fisher information $F_Q$ is the maximum Fisher information attainable over all possible measurements, so that
$
F_C\leq F_Q.
$
The decoding operation therefore preserves the quantum Fisher information while redistributing the encoded information among measurement observables, allowing a realistic measurement to recover a larger fraction of the available $F_Q$. The information remains distributed across the full many-body probability distribution even when Gaussian squeezing disappears~\cite{Str14b}}

The discussion above highlights an important distinction between metrological usefulness and ease of readout. In principle, the available information can be recovered through full distribution reconstruction and Bayesian inference~\cite{Str14b}. In practice, however, such approaches \revision{may be} challenging in precision sensing applications. This observation motivates the search for measurement strategies that retain access to the available Fisher information while relying on simpler observables and realistic detection capabilities.

\subsection{Imperfect detection and readout amplification}

A second limitation arises even when the relevant observable is known. Finite detection resolution can erase the metrological advantage of an entangled state, particularly when the encoded information is stored in \revision{fine features} of the output distribution~\cite{Dav16,Hos16b}. In this regime, the challenge is no longer generating quantum correlations but extracting them experimentally.  

The central idea behind nonlinear readout is simple: if the detector cannot resolve the microscopic structure of the output state, the protocol should not require it to do so. Instead, a controlled many-body interaction is applied after phase encoding to transform the information into a more accessible observable. The readout does not create Fisher information. It changes how that information is represented at the time of measurement.

An important consequence is that nonlinear readout and entanglement generation play distinct roles. If the dominant limitation is classical detection noise, a nonlinear readout can already improve the measurement of an initially coherent spin state. This is the principle underlying quantum phase amplification~\cite{Hos16b}: a nonlinear interaction amplifies a small encoded phase into a larger collective signal that can be resolved by a noisy detector. Entanglement becomes necessary only when the intrinsic sensitivity of the probe is intended to surpass the SQL~\cite{Hos16b}.

\begin{figure}[ht!]
    \centering
    \includegraphics[width=0.75\linewidth]{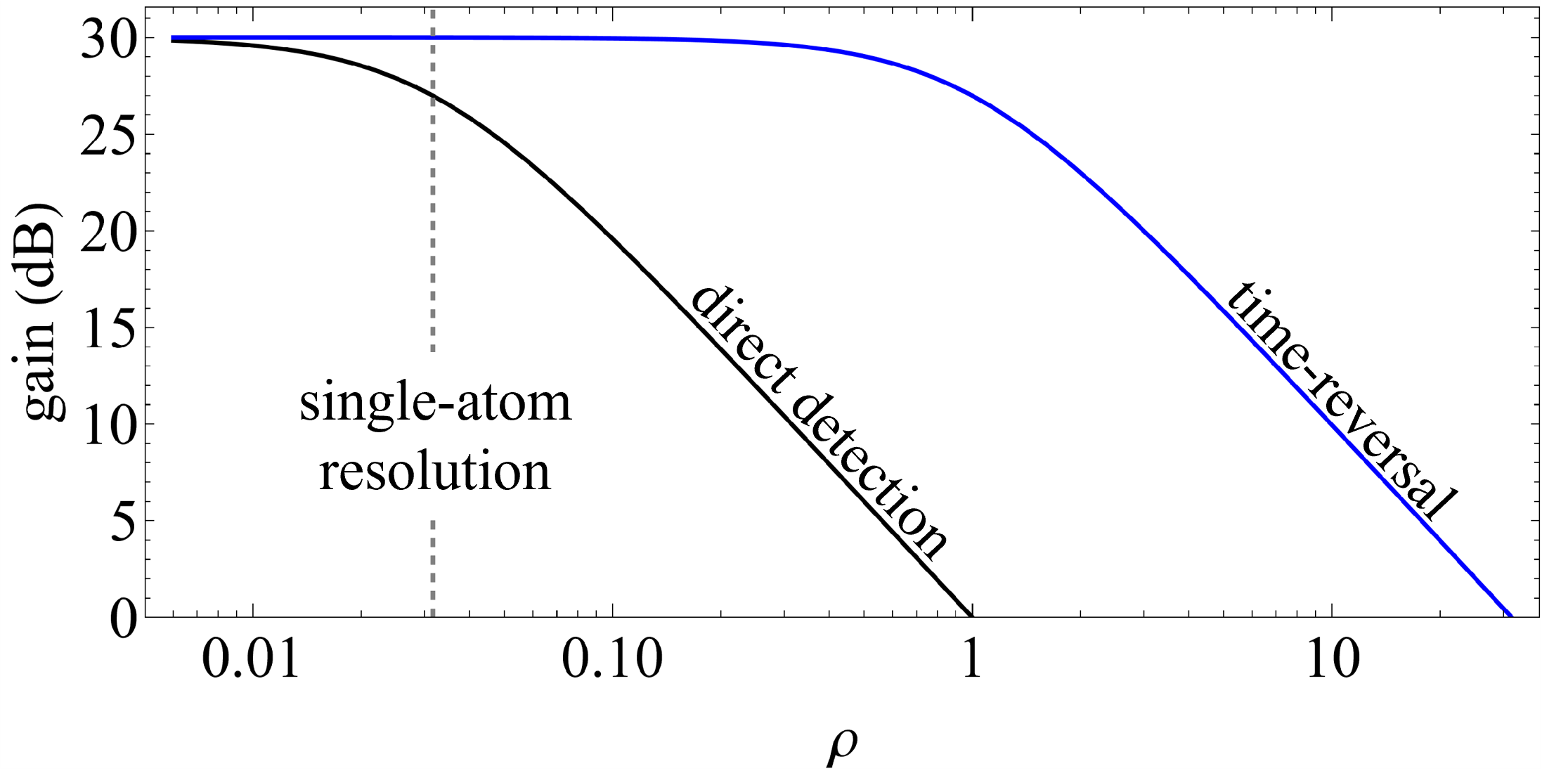}
    \caption{Comparison of the best achievable \revision{metrological} gain as a function of the normalized measurement noise $\rho$ in the case of $N=1000$ atoms. The black line shows the case of direct detection, while the blue line is the ideal time-reversal protocol where the final state is a CSS.}
    \label{fig:gain_vs_meas_res}
\end{figure}

The twist--untwist proposal of Davis \emph{et al.}~\cite{Dav16} makes this point particularly clear. An entangling evolution first generates a nonclassical many-body state. After phase encoding, an inverse interaction amplifies the signal while restoring the quantum noise to that of a coherent spin state. If the measurement adds \revision{independent} Gaussian noise  $\Delta S_{\rm meas}=\rho\,\Delta S_{\rm CSS}$, \revision{with $\rho$ the normalized measurement noise,} the resulting phase sensitivity becomes
\begin{equation}
    \Delta \phi = \sqrt{1+\rho^2}\,\Delta \phi_{\min},\label{eq:measurementNoiseInf}
\end{equation}
showing that near-optimal sensitivity can be maintained even when the detector only marginally resolves projection noise~\cite{Dav16}, as represented in Fig.~\ref{fig:gain_vs_meas_res}. The inverse interaction does not generate additional metrological information. Its role is to convert information encoded in a complex many-body state into a signal that remains accessible under realistic detection conditions. 

A complementary strategy was demonstrated experimentally by Hosten \emph{et al.} under the name of quantum phase magnification~\cite{Hos16b}. Using the one-axis-twisting Hamiltonian $\hat H=\hbar\chi \hat S_z^2$, the nonlinear evolution generates the mapping

\begin{equation}
    \hat S_z(t) = \hat S_z(0),
    \qquad
    \hat S_y(t) \revision{\simeq}  \hat S_y(0) + m \hat S_z(0),
\end{equation}

where $m$ is the signal magnification factor. For a constant shearing rate,
$
m = N\chi t.
$ 

A small phase or population signal initially encoded in $S_z$ is thereby converted into a larger displacement in $S_y$, making it easier to detect. For squeezed input states, the protocol introduces a small pre-rotation $\theta$ that aligns the amplified signal with the measurement basis while partially refocusing the antisqueezed quadrature. Choosing $m\simeq 1/\theta$ suppresses the contribution of the antisqueezed noise while preserving the signal amplification~\cite{Hos16b}. As a result, the protocol amplifies the encoded signal without fully inheriting the large fluctuations associated with the conjugate quadrature.

The physical mechanism differs from the twist--untwist protocols discussed above. The nonlinear evolution is not reversed through a change in the sign of the shearing Hamiltonian. Instead, the shearing dynamics and subsequent spin rotations are engineered so that the encoded phase is converted into a larger collective displacement before detection~\cite{Hos16b}. Experimentally, this enabled squeezed-state metrology approximately $8$~dB below the standard quantum limit using a detector whose technical noise floor was about $10$~dB above the standard quantum limit~\cite{Hos16b}. The result demonstrates that the practical advantages of nonlinear readout do not rely on a literal time reversal of the many-body dynamics.

Davis \emph{et al.} ~\cite{Dav16} and Hosten \emph{et al.}~\cite{Hos16b} highlight a common challenge in quantum-enhanced metrology: the information available in an entangled state is often encoded on scales that are inaccessible to realistic detectors. The role of the post-encoding interaction is therefore not to generate additional sensitivity, but to make the existing sensitivity experimentally accessible. In the twist--untwist protocol, this is achieved through an approximate reversal of the entangling dynamics. In phase magnification, it is achieved through nonlinear amplification followed by a suitable spin rotation. Despite these differences, both approaches exploit the same principle: a controlled many-body evolution is used to transform microscopic phase information into a robust collective signal before measurement. This idea lies at the heart of nonlinear readout protocols and provides much of the motivation for the time-reversal-based approaches discussed throughout this review.

\begin{figure}[t!]
    \centering
    \includegraphics[width=0.79\linewidth]{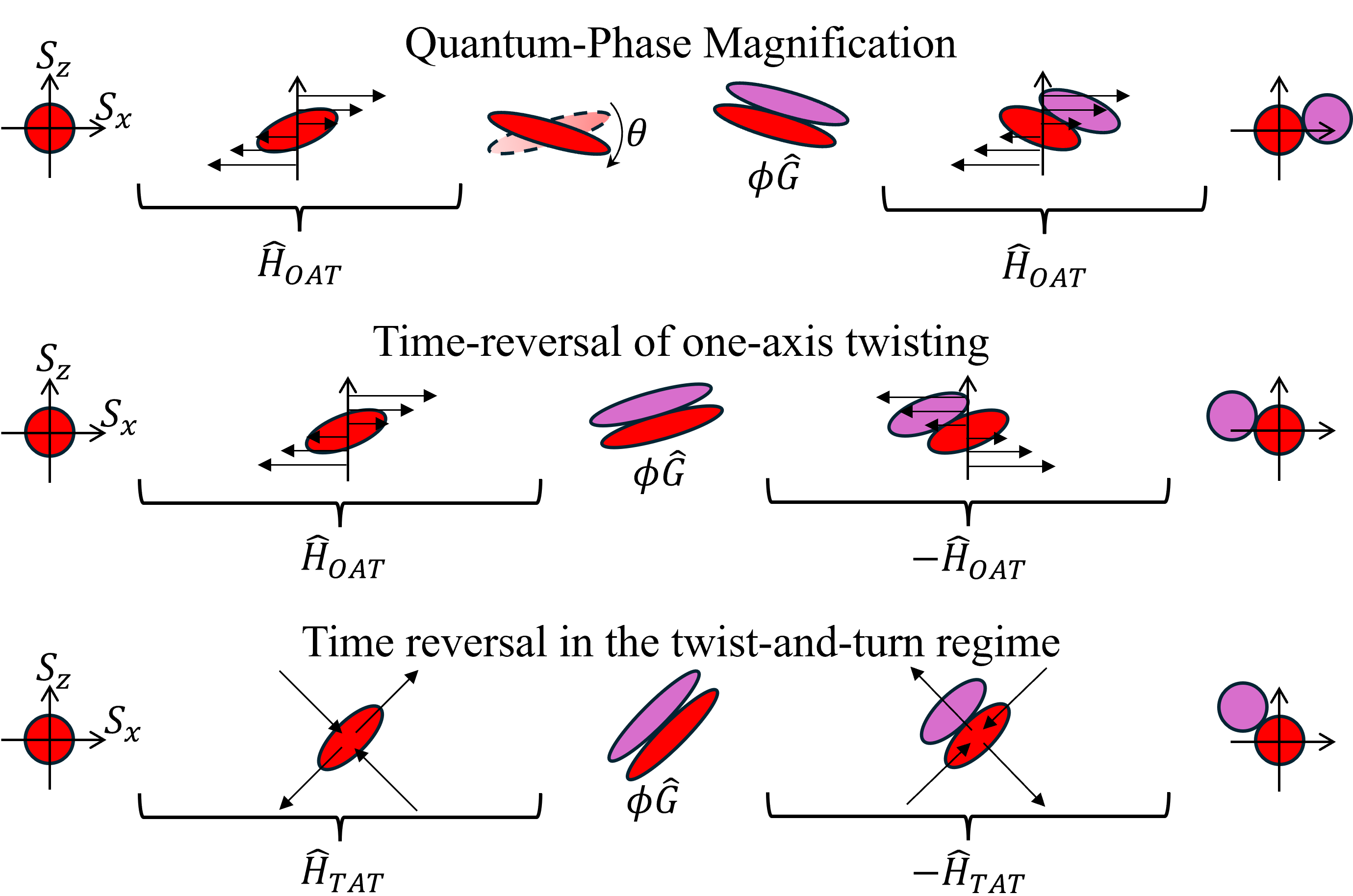}
    \caption{\textbf{Conceptual comparison of quantum phase magnification and time-reversal protocols in collective-spin phase space.} An initially coherent spin state is transformed by nonlinear dynamics before a small phase $\phi \hat G$ is encoded. \revision{The arrows indicate the local evolution direction and rate in phase space at each point.} Top: Quantum phase magnification uses two one-axis-twisting (OAT) evolutions separated by a rotation ($\theta$), amplifying the encoded phase without reversing the Hamiltonian. Middle: The SATIN protocol generates an entangled state through OAT and subsequently untwists it by reversing the interaction -$\hat H_\mathrm{OAT}$, mapping the encoded phase onto a large collective-spin displacement. Bottom: The same principle is applied in the twist-and-turn regime, where reversal of the nonlinear dynamics enables interaction-based readout of more rapidly generated non-Gaussian states.}
    \label{fig:Comparison}
\end{figure}

\section{Practical implementations in cavity QED}

The experimental question is not merely whether a protocol contains a backward step, but which part of the dynamics is inverted and for what metrological purpose. In cavity-based spin systems, this distinction becomes particularly transparent with phase magnification and spin-boson echoes. The microscopic implementations differ, yet a common structure persists: a collective degree of freedom, a bosonic mediator, a controllable nonlinear evolution, and a final measurement whose performance depends on that evolution~\cite{Hos16b,Col21,Li22c,Gil21}.

The experimental landscape remains sufficiently compact that its main developments can be traced through a few representative platforms. One direction uses cavity-mediated interactions as a nonlinear readout, without attempting a literal inversion of the dynamics~\cite{Hos16b}. 
A second implements Hamiltonian reversal in collective-spin cavity systems, first for one-axis twisting and later in the Lipkin--Meshkov--Glick regime~\cite{Col21,Li22c}. Figure \ref{fig:Comparison} summarizes the operating principles of the three cavity-based nonlinear readout protocols discussed in this section. 
A third implementation reaches the same operational structure in a trapped-ion setting, where the bosonic mediator is phononic rather than photonic~\cite{Gil21}. Together, these experiments illustrate how interaction-based readout evolved from a theoretical concept into a practical tool for quantum-enhanced metrology.

\subsection{\revision{Experimental landscape and demonstrated performance}}

\revision{The experimental implementations of interaction-based readout and time reversal span several physical platforms, including cavity-coupled atomic ensembles, Bose--Einstein condensates, and trapped-ion systems. Table~\ref{tab:experimental_IBR} summarizes these realizations in terms of the system size, readout or reversal operation, and experimentally characterized performance. We distinguish between \emph{characterized metrological gain}, which includes experimentally quantified enhancement of a metrological resource or sensitivity, and \emph{demonstrated sensing gain}, which refers to an enhancement directly demonstrated in a complete sensing measurement. This distinction allows the reported results to be compared across experiments that use different sensing variables and performance normalizations. 
}

\revision{The entries in Table~\ref{tab:experimental_IBR} involve different prepared states, sensing variables, and readout strategies. The prepared states and the corresponding interaction dynamics determine the metrological resource available before readout, while the reversal or readout operation converts this resource into an experimentally accessible signal. Detection noise and decoherence therefore play a central role in the experimentally achieved performance. These effects, together with the relevant prepared states and platform-specific limitations, are discussed below for each implementation.}

\begin{table}[h!]
\bgroup
\def\arraystretch{1.5}
\caption{Experimental implementations of interaction-based readout and time-reversal protocols and their achieved performance.}
\label{tab:experimental_IBR}%
\centering
\small
\begin{tabular}{@{}p{0.19\textwidth}p{0.08\textwidth}p{0.27\textwidth}p{0.17\textwidth}p{0.17\textwidth}@{}}
\toprule
Platform & $N$ approx. & Interaction / reversal & Characterized metrological gain & Demonstrated sensing gain \\
\midrule
$^{87}$Rb cavity~\cite{Hos16b}
& $5\times10^5$
& OAT; interaction-based readout
& $\sim8$~dB
& --
\\
$^{171}$Yb cavity~\cite{Col21}
& $350$
& OAT; optical-detuning reversal
& $12.8\pm0.9$~dB 
& $11.8\pm0.5$~dB
\\
$^{171}$Yb cavity~\cite{Li22c}
& $200$
& $H=\chi S_z^2-\Omega S_x$; twist-and-turn reversal
& $6.8\pm0.4$~dB\newline 
& --
\\
Trapped ions~\cite{Gil21}
& $150$
& Spin--boson coupling; $\pi$-pulse reversal
& $8.8\pm0.4$~dB\newline (displacement)
& $4.0\pm0.5$~dB\newline(E-field SQL)
\\
Spinor BEC~\cite{Lin16}
& $\sim2.8$
& Nonlinear spinor dynamics; $H\rightarrow-H$
& $3.7\pm1.6$~dB
& --
\\
$^{87}$Rb BEC~\cite{Mao22}
& $26{,}400$
& Spin-nematic interaction; echoed reversal
& $21.6\pm0.5$~dB
& $16.6\pm1.1$~dB
\\
\bottomrule
\end{tabular}
\egroup
\end{table}

\subsection{Quantum phase magnification}

Hosten \emph{et al.} demonstrated that nonlinear readout does not require a literal reversal of the underlying many-body dynamics in order to recover metrological information that would otherwise be hidden by the final detection stage~\cite{Hos16b}. The experiment was carried out with an ensemble of $^{87}$Rb atoms coupled dispersively to an optical cavity, where the collective spin dynamics are described by the one-axis-twisting Hamiltonian.
$
\hat H_\mathrm{OAT} = \hbar \chi \hat{S}_z^2,
$
with $\hat S_z$ the population imbalance between the two clock states. The cavity generates this interaction through a dispersive feedback mechanism: fluctuations in $S_z$ shift the cavity resonance frequency modifying the intracavity photon number and the resulting ac Stark shift, which in turn produces an effective nonlinear interaction proportional to $S_z^2$~\cite{Hos16b}.

Around a state polarized along $x$, it is convenient to introduce the Holstein--Primakoff description with canonical fluctuation variables $X=S_y/\sqrt{S}$ and $P=S_z/\sqrt{S}$, for which the one-axis-twisting evolution acts as a shear in phase-space. In this picture, the phase signal is first converted into a small population difference, corresponding to a displacement along $S_z$. The state is then subjected to a carefully chosen rotation about the $x$ axis before undergoing a second nonlinear evolution under the same Hamiltonian~\cite{Hos16b}. 

The protocol therefore keeps the Hamiltonian unchanged. It does not attempt $\chi\rightarrow -\chi$. Instead, the intermediate rotation angle $\theta$ is tuned such that, within the Holstein-Primakoff approximation its combination with the subsequent one-axis-twisting evolution \revision{mimics an effective time reversal of the OAT dynamics in the relevant phase-space region}. 
The state evolution is described by 
$$
|\psi_{\phi}\rangle = e^{-\frac{i}{\hbar} \chi \hat{S}_z^2 t} e^{-\frac{i}{\hbar}\theta \hat{S}_x} e^{-\frac{i}{\hbar}\phi \hat{S}_y} e^{-\frac{i}{\hbar}\chi \hat{S}_z^2 t}|x\rangle,
$$
where $|x\rangle$ denotes the initial coherent spin state oriented along the $x$-axis. With this notation, in the Holstein-Primakoff approximation, Hosten et al. constructed $\theta$ so that 
$$
|\psi_{\phi}\rangle \approx e^{\frac{i}{\hbar} \chi \hat{S}_z^2 t} e^{-\frac{i}{\hbar}\phi \hat{S}_y} e^{-\frac{i}{\hbar} \chi \hat{S}_z^2 t}|x\rangle.
$$

Written this way, the logic is close in spirit to time-reversal readout but not identical to it. The second nonlinear stage partially removes the observable imprint of the originally generated entanglement not by undoing the many-body evolution term by term, but by driving the measured quadrature back toward a simple population signal. For squeezed inputs, ~\cite{Hos16b} adds a separate small pre-magnification rotation $\theta$ that refocus the anti-squeezed noise at $m\simeq 1/\theta$. That refinement is useful experimentally, but it should not be confused with the main readout step itself.

This experiment is therefore better read as a cavity implementation of interaction-based readout than as time reversal in the strict sense. The same physical resource that produces the nonlinear evolution is used to make the encoded information accessible, but no Hamiltonian inversion is required. That distinction, although easy to blur at the level of pulse diagrams, becomes important once one compares later cavity experiments that do rely on a true reversal of the effective many-body dynamics.

\paragraph{Limitations}
In the quantum magnification protocol \cite{Hos16b}, the unfavorable growth of the required magnification with increasing squeezing is avoided, but only if the initial pre-rotation and the nonlinear \revision{evolution} are properly matched, with optimal refocusing near $m\simeq 1/\theta$. The protocol is therefore sensitive to rotation inaccuracies, since away from this condition, residual anti-squeezed noise reappears in the final readout. It is also fundamentally limited to the small-angle planar regime, i.e., in the Holstein-Primakoff approximation: the \revision{metrological} gain cannot be increased arbitrarily without eventually encountering Bloch-sphere curvature and state wrapping. In practice, the readout in \cite{Hos16b} remains constrained by technical fluorescence-detection noise, with smaller deviations from ideal behavior arising from slow squeezing drifts and residual atom-cavity coupling inhomogeneity.

\subsection{Collective-spin time reversal in the Yb cavity platform}

The direct cavity implementation of Hamiltonian-level time reversal was realized in the $^{171}$Yb platform developed by Vuleti\'{c}'s group. In ~\cite{Col21}, an ensemble trapped inside an optical cavity evolves under an effective one-axis-twisting Hamiltonian,
$
H =\revision{\hbar} \chi S_z^2,
$
with a normalized twisting strength
$
\tilde{Q}=\sqrt{N}\chi\tau.
$

\begin{figure}
    \centering
    \includegraphics[width=0.7\linewidth]{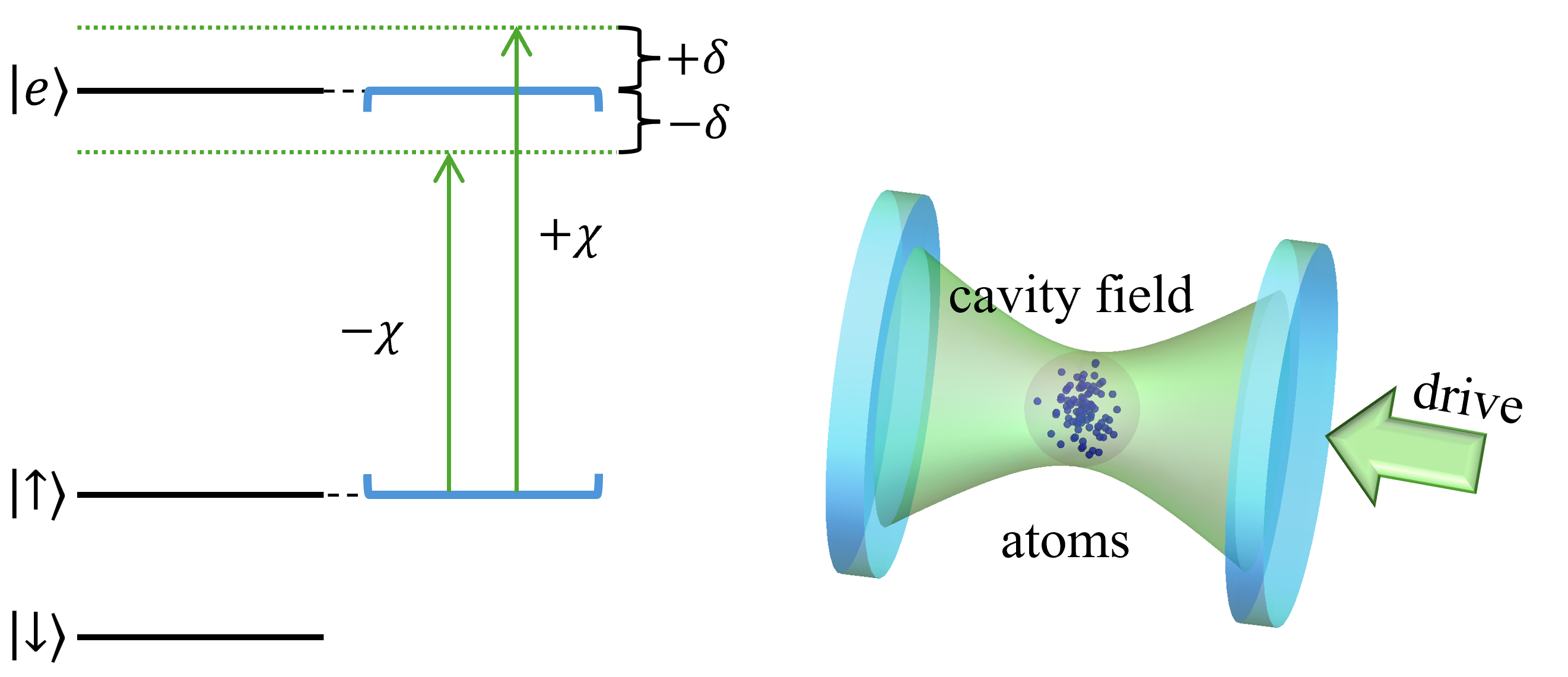}
    \caption{\textbf{Experimental realization of the SATIN protocol in the cavity-QED platform.} In ~\cite{Col21} a collective spin of $^{171}$Yb atoms interacts dispersively with a single optical cavity mode. Driving one dressed atom-cavity resonance produces a cavity-mediated one-axis-twisting interaction $H=\chi S_z^2$, generating a non-Gaussian entangled state. After phase encoding, the probe frequency is switched to the opposite dressed resonance, reversing the sign of the cavity feedback and implementing $\chi\rightarrow -\chi$. This time-reversed evolution untwists the state and maps the encoded phase onto an amplified collective-spin displacement, allowing near-Heisenberg-limited sensitivity with detection noise at the coherent-spin-state level.}   
   \label{fig:Yb}
\end{figure}

Changing the optical detuning reverses the sign of $\chi$, so the post-encoding step does correspond to a genuine untwisting operation~\cite{Li21b,Col21}, as shown in Figure~\ref{fig:Yb}. The protocol is described by the sequence

\begin{equation}
U(-\tilde{Q})\,R_y(\phi)\,U(\tilde{Q}),
\end{equation}

or, more general\revision{ly} in the full Ramsey implementation, by an entangling evolution followed by phase encoding and evolution under the reversed Hamiltonian~\cite{Col21}. The encoded phase is converted into an amplified collective-spin displacement,

\begin{equation}
\langle \hat S_y\rangle \simeq m\phi S,
\end{equation}

with an amplification factor that depends on the shearing strength. The analytical model gives, to leading order,

\begin{equation}
m(\tilde{Q}) \approx C_{\mathrm{sc}}(\tilde{Q})\,(N\revision{-1})\sin\!\left(\frac{\tilde{Q}}{\sqrt{N}}\right)\cos^{N\revision{-2}}\!\left(\frac{\tilde{Q}}{\sqrt{N}}\right),
\end{equation}

where $C_{\mathrm{sc}}$ accounts for scattering-induced contrast loss~\cite{Col21}. In that form, the metrological role of the inverse step is explicit: it maps a phase that had become inaccessible in a non-Gaussian state back onto a directly measurable collective observable.

Reference ~\cite{Li22c} extends the same apparatus into a different dynamical regime rather than merely refining the same protocol. The effective Hamiltonian becomes
\begin{equation}
\hat H = \revision{\hbar \large(}\chi \hat S_z^2 - \Omega \hat S_x\revision{\large)},
\end{equation}
so the reversal must act on both the nonlinear term and the collective rotation. Near the unstable fixed point, the dynamics acquire the twist-and-turn character associated with rapid scrambling. The relevant control is the ratio
$
\Omega/ S\chi,
$
for which the short-time dynamics reduce to an effective two-axis-twisting Hamiltonian,
\begin{equation}
\hat H_{\mathrm{TAT}} = \revision{\hbar} \chi\left(\hat S_z^2-\hat S_y^2\right),
\end{equation}
within the Holstein--Primakoff description~\cite{Li22c}. Experimentally, the reversal is implemented by switching the cavity drive to change $\chi\rightarrow -\chi$ and reversing the phase of the radio-frequency drive to obtain $\Omega\rightarrow -\Omega$~\cite{Li22c}. The metrological signatures are then tracked through the growth of signal amplification and system noise, yielding $6.8(4)$~dB \revision{metrological} gain beyond the standard quantum limit.

Taken together, Refs.~\cite{Col21,Li22c} define a single experimental thread. The first establishes that cavity-mediated untwisting can serve as a practical readout tool in a full interferometer. The second shows that the same control techniques remain effective in a qualitatively different dynamical regime, where a transverse drive modifies the phase-space structure and the dynamics move beyond pure $S_z^2$ evolution. 

A recent extension of this line of work appears in global-phase spectroscopy on an optical clock transition~\cite{Zap25}. There, time-reversal readout is combined with rotary echo protection against inhomogeneous coupling. Beyond the specific implementation, the significance of this result is that it carries the cavity-based time reversal toolbox into optical transition spectroscopy, further broadening the range of precision measurements that can benefit from interaction-based readout.

\paragraph{Limitations}
\revision{The scalability of SATIN-like protocols is ultimately limited by classical noise and control and not by atom number and cooperativity $\eta$. Although Heisenberg scaling can persist over all atom numbers, atom loss and other decoherence mechanisms can restore SQL scaling at sufficiently large $N$~\cite{demkowicz2012elusive}. In~\cite{Col21}, finite-vacuum-induced atom loss of $\gamma_\mathrm{Loss}=0.1~\mathrm{s^{-1}}$, should induce this crossover near $N\approx4\times10^4$ for a Ramsey time of $3.46$~ms. Increasing the single-atom cooperativity can reduce dissipation and extend the regime of enhanced scaling, provided that the twisting and untwisting operations remain sufficiently well controlled.}

A \revision{common} control limitation in the time-reversal-based readout schemes of \cite{Col21, Li22c} is that the metrological advantage depends on accurate inversion of the nonlinear dynamics: the untwisting step must closely match the initial twisting strength and timing, so imperfections lead to incomplete refocusing and residual distortion of the readout signal. In the bare one-axis-twisting case \cite{Col21}, the main practical limitations are dissipation and technical imperfections during cavity-mediated twisting, especially photon-scattering-induced contrast loss, excess broadening from light carrying away spin information, and sensitivity to control of the shearing and unshearing pulses. The twist-and-turn experiment \cite{Li22c} is further constrained by the demands of the Hamiltonian architecture and its complexity, including precise sign reversal of the interaction and a higher susceptibility to spontaneous emission. Thus, the performance of direct cavity implementation of time-reversal protocols is constrained by decoherence and dissipation during the interaction, including photon scattering, contrast loss, and excess broadening from information leaked into the light field; it is worth noting that cavities with higher single-atom cooperativities mitigate these losses further.

\subsection{Trapped ions as a spin–boson realization of cavity-QED protocols}

The trapped-ion experiment~\cite{Gil21} realizes the same operational structure discussed above, but in a different physical setting. Although no optical cavity is involved, the ingredients are closely analogous: a collective spin, a quantized bosonic mode, a tunable spin-boson coupling, and an echo step that maps a weak signal onto a robust collective observable. In the ion crystal, the bosonic mode is the center-of-mass phonon and the coupling is produced by a spin-dependent optical dipole force rather than by intracavity photons.

In the interaction picture used in~\cite{Gil21}, the dominant coupling takes the form

\begin{equation}
\hat{H}_{\mathrm{ODF}} = \frac{\hbar g}{\sqrt{N}}\left(a+a^{\dagger}\right)S_z + \revision{\hbar}\,\delta a^{\dagger}a,
\end{equation}

where $a$ and $a^{\dagger}$ act on the center-of-mass mode, $S_z$ is the collective spin, and $\delta$ is the detuning from the motional resonance. The sensing protocol can be written schematically as

\begin{equation}
|\psi_f\rangle = e^{\revision{\frac{i}{\hbar}}\hat H_{\mathrm{ODF}} \tau}U_{\beta}e^{-\revision{\frac{i}{\hbar}}\hat H_{\mathrm{ODF}}\tau}|\psi(0)\rangle,
\end{equation}

with $U_{\beta}$ a small coherent displacement of the center-of-mass motional mode to be detected~\cite{Gil21}. This differs from the optical-cavity spin-sensing protocols discussed above, where the weak signal is encoded as a small collective-spin rotation, since here the perturbation is first imprinted on the bosonic mediator itself and only converted into a spin observable by the echo.
Operationally, the reversal is implemented through a microwave $\pi$ pulse inserted between two equal interaction windows, effectively reversing the sign of the spin-boson coupling during the second half of the sequence.

\revision{For resonant coupling, $\delta=0$ and a displacement amplitude $\beta$, the echo maps the motional displacement onto a collective spin rotation,}

\begin{equation}
|\psi_f\rangle = e^{i\phi \hat S_z}U_{\beta}|\psi(0)\rangle,
\qquad
\phi = \frac{2g\tau\beta}{\sqrt{N}}.
\end{equation}

Therefore, the motional signal is read out through the spin degree of freedom rather than from direct oscillator detection~\cite{Gil21}. The reported performance, $8.8 \pm 0.4$~dB below the standard quantum limit for displacement sensing and $240 \pm 10$~nV~m$^{-1}$ in $1$~s for electric-field sensing, illustrates the practical power of this approach.

A useful precursor in the trapped-ion setting is the single-ion experiment of \cite{burd2019}, which demonstrated quantum amplification of a weak motional displacement by placing the signal between a squeezing operation and its inverse. The sensed quantity is again a small displacement of a harmonic mode, but the protocol acts directly on the oscillator rather than on a collective spin-boson system: the ion's radial motional mode is first squeezed, then weakly displaced along the squeezed quadrature, and finally anti-squeezed so that the displacement is magnified before readout through the internal-state qubit. This realizes the same broad interaction-based-readout logic of coherently amplifying a small signal before measurement, but through reversible single-mode bosonic dynamics rather than the many-body echo used in~\cite{Gil21}.

From the perspective of this review, the importance of~\cite{Gil21}~and~\cite{burd2019} lies in showing that the essential ingredients of time-reversal metrology are not unique to cavity QED. Once a controllable spin-boson interaction is available, the same echo logic can be implemented in a different physical architecture. 

It is useful in this regard to compare ~\cite{Gil21, burd2019} with ~\cite{Fra23}, where finite-range interactions on an optical transition produce entanglement-enhanced sensing in trapped-ion chains without employing the same echo-based reversal of a collective spin-boson map. The comparison highlights what is distinctive about~\cite{Gil21}: not the trapped-ion platform itself, but the ability to reproduce the operative structure of a cavity-style time-reversal protocol.

\paragraph{Limitations}
In trapped-ion motional amplification protocols, the metrological gain remains contingent on precise phase control and accurate reversal or matching of the amplification dynamics, so mistuning degrades the final signal rather than cleanly amplifying it \cite{Gil21, burd2019}. Their operation is also restricted to the appropriate small-excursion regime. In \cite{Gil21}, this is due to the collective-spin mapping used in the large-$N$ Holstein-Primakoff treatment, while in \cite{burd2019}, it is due to the linear-response and Lamb-Dicke conditions required for faithful sideband-based readout. In practice, performance is further limited by motional-frequency noise, residual spin-phonon entanglement, and spin decoherence from light scattering in the many-body echo setting, as well as motional heating, dephasing, trap anharmonicity, and projection-noise-limited qubit readout in the single-oscillator parametric-amplification setting.

\section{Outlook and perspectives}

The developments reviewed above suggest that the role of many-body interactions in quantum metrology is changing. Historically, the emphasis was on generating entanglement and quantifying its metrological usefulness through spin squeezing or related measures~\cite{pezze2018quantum,kaubruegger2025progress}. More recent work has shown that state preparation is only part of the problem. As experiments move into regimes involving strongly non-Gaussian states, oversqueezed states, and more complex forms of multipartite entanglement, the extraction of the encoded information becomes equally important~\cite{Col21}.

From this perspective, interaction-based readout and time-reversal protocols are not merely techniques for improving measurement sensitivity. They provide a framework for controlling how information flows through a many-body system before it is measured. The same interactions that generate entanglement can also be used to process, amplify, and decode it. This shift naturally raises broader questions about the types of quantum states that can be used for metrology, the role of reversible many-body dynamics as a resource, and the range of experimental platforms in which these ideas can be implemented.

\subsection{Beyond spin squeezing: decoding complex quantum states}

The first generation of quantum-enhanced metrology focused largely on spin-squeezed states~\cite{ma2011quantum}. These states provided a natural starting point because they could be generated with controllable interactions and characterized through relatively simple observables such as collective spin variances. However, nonlinear many-body evolution naturally produces states that extend far beyond the Gaussian regime. Oversqueezed states~\cite{Str14b,Col21,Li22c}, Dicke states~\cite{zhang2014quantum,lucke2014detecting,lucke2011twin}, GHZ states~\cite{pezze2018quantum,cao2024multi,kessler2014heisenberg}, \revision{spin cat states~\cite{Hua18a} (EP)}, and other forms of multipartite entanglement can possess substantially larger quantum Fisher information than conventional squeezed states, while exhibiting probability distributions that are increasingly difficult to characterize and measure~\cite{pezze2018quantum,gessner2019metrological}.

Historically, this complexity was often viewed as a limitation. Although highly entangled states could in principle offer superior sensitivity, extracting the encoded information typically required measurements that were experimentally inaccessible or prohibitively demanding. One of the central lessons emerging from interaction-based readout and time-reversal protocols is that the usefulness of a quantum state cannot be assessed independently of the measurement strategy. A state that appears impractical under direct detection may become metrologically valuable when combined with an appropriate decoding operation. This perspective shifts the emphasis from entanglement generation alone to the joint design of state preparation and readout.

Looking forward, increasingly complex many-body states will likely require increasingly sophisticated decoding strategies. Recent proposals based on scrambling dynamics, generalized echo protocols, and engineered decoding operations point in this direction, extending the logic of interaction-based readout beyond the one-axis-twisting paradigm that originally motivated SATIN and related protocols~\cite{Li22c,Car26,bringewatt2026butterfly,liu2026echoed,shao2026enhanced,Hua18b,Ma24}. Cavity QED is particularly well suited to explore these ideas because the same interactions used to generate entanglement can also be used to process, amplify, and decode it. As control over many-body dynamics continues to improve, the distinction between state preparation and measurement may become progressively less meaningful, with both becoming integrated components of a single metrological protocol.

\subsection{Reversible many-body dynamics as a resource}

One of the most significant conceptual developments emerging from time-reversal metrology is the realization that reversibility itself can become a useful resource. In the examples discussed throughout this review, the same interactions that generate entanglement are later used to amplify signals, decode complex quantum states, or reverse the effects of many-body evolution. From this perspective, the ability to control many-body dynamics is not merely a technical requirement for quantum-enhanced sensing. It becomes part of the sensing protocol itself.
 
This viewpoint establishes natural connections between quantum metrology and broader questions in many-body physics. Concepts originally associated with Loschmidt echoes, scrambling, quantum chaos, and the emergence of irreversibility acquire direct operational significance when embedded within a measurement protocol. Conversely, metrological platforms provide experimentally accessible settings in which the reversibility of complex quantum dynamics can be studied quantitatively. A particularly important early example came from trapped ions, where time reversal of an engineered long-range Ising evolution was used not for metrological readout but to measure out-of-time-order correlators and multiple-quantum-coherence spectra~\cite{garttner2017measuring}. That experiment showed that reversing the sign of the many-body dynamics can map otherwise inaccessible information about operator spreading and correlation growth onto directly measurable observables, including the detection of up to eight-body correlations in a system of more than one hundred ions. In this sense, the same control paradigm that later became central to interaction-based readout already appeared as a tool for diagnosing scrambling, reversibility, and the complexity of many-body dynamics, making clear at an early stage that time-reversal protocols have applications well beyond sensing.

Recent experiments based on one-axis twisting, scrambling dynamics, twist-and-turn dynamics, and Lipkin--Meshkov--Glick Hamiltonians suggest that the metrological value of a quantum state may depend not only on the entanglement it contains, but also on the ability to manipulate its subsequent evolution~\cite{Col21,Li22c}.

Looking forward, reversible many-body dynamics may become a resource that can be engineered and optimized alongside more familiar quantities such as coherence time, particle number, and interaction strength. Future quantum sensors may be characterized not only by the amount of entanglement they generate, but also by their ability to process, decode, and recover information stored in increasingly complex many-body states. In this sense, the control of reversibility may emerge as a fundamental ingredient of quantum-enhanced metrology.

An important challenge moving forward will be to develop decoding protocols that remain effective under realistic experimental conditions. While the fundamental limits imposed by decoherence and noise on quantum-enhanced metrology are now well understood~\cite{escher2011general,demkowicz2012elusive}, interaction-based readout introduces additional opportunities for optimizing the decoding stage itself. Rather than seeking exact Hamiltonian reversal, future protocols will likely combine engineered many-body dynamics with realistic noise models to maximize the experimentally accessible information~\cite{Nol17,Hai18,Sch19,Liu24}.

\subsection{Emerging platforms and future opportunities}

Although many of the experimental realizations discussed in this review originated in cavity QED, the underlying ideas are considerably more general. Time-reversal protocols and interaction-based readout have already been demonstrated in spinor Bose--Einstein condensates~\cite{Lin16} and trapped-ion systems~\cite{burd2019,Gil21}, showing that the essential ingredients are not tied to a specific hardware architecture. In spinor condensates, nonlinear spin-changing collisions generate coherent pair-production dynamics whose effective sign can be reversed through controlled phase evolution~\cite{Lin16}. In trapped ions, collective phonon modes mediate long-range spin-spin interactions that closely parallel the role played by the cavity field in cavity QED, enabling echo-based decoding protocols through reversible spin-boson dynamics~\cite{burd2019,Gil21}. More generally, the common requirement is not a particular physical platform, but the ability to generate, control, and approximately reverse a collective many-body evolution.

\revision{A particularly promising direction is the extension of these techniques to new platforms in which collective spins interact through bosonic modes. Cavity photons, collective excitations, and phonons provide distinct physical realizations of this general interaction structure, suggesting that the concepts developed here can be transferred to a broader class of quantum system}
Several emerging platforms offer new opportunities along these lines. Neutral-atom tweezer arrays combine programmable Rydberg interactions with single-particle control, enabling the deterministic preparation of increasingly complex entangled states and precise manipulation of many-body dynamics~\cite{bernien2017probing,browaeys2020many,manetsch2025tweezer,nakamura2024hybrid,kaufman2021quantum,finkelstein2024universal,bornet2023scalable}. These capabilities have already transformed quantum simulation and quantum information science and may ultimately provide a natural setting for generalized interaction-based readout, scrambling-assisted metrology, and optimal decoding of highly entangled states~\cite{cao2024multi,eckner2023realizing,bornet2023scalable}. At the same time, multimode cavity QED and engineered long-range spin Hamiltonians continue to expand the range of collective dynamics that can be realized experimentally, opening new opportunities to explore reversible quantum dynamics beyond the one-axis-twisting paradigm~\cite{marsh2025multimode,shaw2026cavity,vaidya2018tunable,mottl2012roton,bonifacio2024laser,orsi2024cavity}.

Another promising direction is provided by dipolar quantum gases based on highly magnetic atoms such as dysprosium and erbium~\cite{chomaz2023dipolar}. In these systems, long-range magnetic dipole-dipole interactions naturally generate collective nonlinear dynamics that differ qualitatively from the contact interactions of conventional Bose--Einstein condensates~\cite{chomaz2023dipolar}. Recent advances have demonstrated spin squeezing~\cite{douglas2025}, controllable many-body interactions, and high-fidelity coherent control in dipolar gases~\cite{grun2024}, suggesting that they may eventually support interaction-based readout and reversible metrological protocols based on intrinsically long-range interactions~\cite{douglas2025,grun2024}.

Solid-state systems represent a complementary direction toward scalable quantum sensing. Ensembles of interacting spins, color centers~\cite{gao2025signal}, superconducting qubit architectures~\cite{hu2026quantum,kobrin2024universal,blais2021circuit,blais2020quantum}, and other engineered quantum platforms combine collective quantum control with mature fabrication technologies and scalable architectures~\cite{degen2017quantum}. Recent theoretical proposals indicate that interaction-based readout, quantum phase amplification, and reversible many-body dynamics can also be implemented in these systems, extending concepts originally developed in atomic physics into the solid-state domain~\cite{gao2025signal}. Achieving the required coherence and Hamiltonian control remains challenging, but the potential impact on quantum sensing and quantum information processing is substantial.

Ultracold molecules provide another particularly promising opportunity. Their rich internal structure, strong electric dipole moments, long coherence times, and exceptional sensitivity to weak interactions make them attractive candidates for precision measurements and tests of fundamental physics~\cite{demille2024quantum,ye2024essay,kaufman2021quantum,holland2026creating}. The combination of cavity-mediated interactions, nonlinear readout, and molecular spectroscopy may enable new approaches to searches for symmetry-violating effects, precision frequency metrology, and other signatures of physics beyond the Standard Model.

Taken together, these developments suggest that reversible quantum metrology is evolving into a platform-independent framework rather than a collection of specialized protocols. The central challenge is no longer simply the generation of entanglement, but the ability to control, manipulate, and recover information stored in increasingly complex many-body systems. As new experimental platforms continue to mature, interaction-based readout and reversible many-body dynamics are likely to become increasingly general tools for extracting quantum-enhanced sensitivity across a broad range of quantum technologies.

\section{Conclusion}

Time-reversal and, more generally, dynamics-reversal protocols have evolved from tools for studying reversibility into practical resources for quantum-enhanced metrology. The central lesson emerging from recent work is that many-body interactions need not be viewed solely as a means of generating entanglement. The same dynamics can also be used to process, amplify, and decode the information stored in complex quantum states. From this perspective, SATIN and related protocols are not isolated techniques but part of a broader interaction-based-readout framework in which the final many-body evolution is engineered to make quantum-enhanced sensitivity accessible under realistic experimental conditions. Cavity QED has played a leading role in this development because it combines collective enhancement, tunable interactions, and controllable reversibility within a single platform. As experiments continue to move beyond the Gaussian regime and toward increasingly complex forms of entanglement, the ability to decode quantum information may become as important as the ability to generate it, making reversible many-body dynamics a central ingredient of future quantum sensors.

\bmhead{Acknowledgements}
 
We thank Vladan Vuleti\'c and Zeyang Li for insightful discussions, and Daniel McCarron for critical reading of the manuscript.


\section*{Declarations}

\begin{itemize}
\item Funding: S.C. acknowledges partial support from Office of Naval Research under Award No. N00014-26-1-2085. E.P.-P. acknowledges support from the Office of Naval Research under Award No. N00014-26-1-2211.
\item Author contribution: Simone Colombo and Edwin Pedrozo- Pe\~nafiel contributed equally to the conceptualization, formal analysis, writing of the original draft, and review and editing of the manuscript.
\item Data availability: not applicable.
\item Ethical approval: not applicable.
\item Consent to Participate: not applicable.
\item Consent to Publish: not applicable.
\end{itemize}




\bibliography{bibliography_timeReversal}

\end{document}